\documentstyle[12pt,epsfig,amsmath]{article}

\begin{document}
\newcommand{\ra}{\rightarrow}
\newcommand{\be}{\begin{equation}}
\newcommand{\ee}{\end{equation}}
\newcommand{\bea}{\begin{eqnarray}}
\newcommand{\eea}{\end{eqnarray}}
\newcommand{\slp}{\not p\,}
\newcommand{\slq}{\not \,q\,}
\newcommand{\slk}{\not k}
\newcommand{\tr}{ {\rm tr} }
\newcommand{\bard}{d \!\!\rule[6.5pt]{4pt}{0.3pt}}

\begin{titlepage} 
\begin{flushright} 
CERN-TH/2000-317 \\
hep--th/0011102
\end{flushright}

\vspace{0.5in}  
\begin{centering} 
{\bf A Novel Mass Hierarchy and Discrete Excitation Spectra 
from Quantum-Fluctuating D-branes}

\vspace{0.2in} 

{\bf G.K. Leontaris}$^{~a}$ and {\bf N.E.~Mavromatos}$^{~b,c}$
\vspace{0.20in}

$^a$ Theoretical Physics Division, Ioannina
University, GR-45110 Ioannina, Greece.\\
$^b$ CERN, Theory Division, 1211 Geneva 23, Switzerland. \\
$^{\,c}$ Theoretical Physics Group, Department of Physics,
King's College London, Strand, London WC2R 2LS, U.K. \\

\end{centering} 

\vspace*{0.2in}
\begin{centering}

{\bf Abstract}

\end{centering}

{\small We elaborate further on a recently proposed scenario 
for 
generating a mass hierarchy through quantum fluctuations 
of a single $D3$ brane, which represents our world 
embedded in a bulk five-dimensional space time. 
In this scenario, the quantum fluctuations of the D3-brane world 
in the bulk direction, quantified to leading order via a `recoil' 
world-sheet logarithmic conformal field theory approach,
result in the dynamical appearance of a 
supersymmetry breaking (obstruction) scale $\alpha$. This may be 
naturally taken to be at the 
TeV range, in order to provide 
a solution to the conventional 
gauge-hierarchy problem.
The bulk spatial direction is characterized by the dynamical 
appearance of an horizon
located at $\pm 1/\alpha$, inside which the positive energy conditions
for the existence of stable matter are satisfied. 
To ensure the correct value of the four-dimensional 
Planck mass, the bulk string scale $M_s$ is naturally found 
to lie at an 
intermediate energy scale of $10^{14}$ GeV.
As an exclusive feature of the 
$D3$-brane quantum fluctuations (`recoil') we find that, 
for any given $M_5$, there is a discrete mass spectrum 
for four-dimensional Kaluza-Klein (KK) modes 
of bulk graviton and/or scalar fields.  
KK modes with masses 
$0 \le m < \sqrt{2}\alpha << M_s$ 
are found to have wavefunctions peaked, and hence localized, 
on the D3 brane at $z=0$.} 

\vspace{0.8in} 

\begin{flushleft}
CERN-TH/2000-317 \\
\end{flushleft} 

\end{titlepage} 
\section{Introduction}

In recent years considerable interest has been concentrated on 
exploring the idea that space-time
is actually $(4+n)$-dimensional, 
with the four-dimensional observable world being a membrane
(Dirichlet  brane~\cite{SR}) of some
string theory living in the $(4+n)$-dimensional bulk 
space-time~\cite{extra}. 
The literature is growing rapidly,
and many interesting models have been considered, some of which 
appear to have falsifiable predictions in the next-generation
accelerators, such as TeVatron and LHC. 
In some of these models, the extra (bulk) dimensions are taken to
be relatively large, compared to the traditional Planck scale,
implying, for instance, a bulk gravitational scale at the range of
a few TeV. 
In the case of compact extra dimensions
there are induced
modifications of the four-dimensional Newton's law, which may
become phenomenologically important for TeV scale
gravity~\cite{floratos99}. 
In this scenaria, the experimental
success of the inverse-square law of
Newton seemed to imply precisely four non-compact
dimensions only. 

However, the work of ref.~\cite{randal99}
has demonstrated that the
situation is completely different 
in cases  where the higher-dimensional
metric was not factorizable, namely the case
where there is
a {\it warp} factor in front of
the  four-dimensional metric which depends on the  coordinates of
the  bulk extra dimensions. According to this approach, our
universe is a static flat domain-wall which, in the simplest case
of five dimensions, separates two regions of five-dimensional Anti
de Sitter (AdS) space-time. In its simplest
version~\cite{randal99}, the scenario is realized by introducing a
positive energy brane at the origin and a negative energy brane at
distance $z$ where our world is located  and where the graviton
amplitude is exponentially suppressed. Modifications to the above
picture with positive energy branes allowing also the possibility
of infinite extra dimensions, multi-brane solutions, and
supergravity embedded versions were considered in the
literature~\cite{noncompact,GW,GRS}. 
Thus, it is worth noticing that the bulk dimensions are not
necessarily compact. The rather  important point of
\cite{randal99}, however, was the demonstration of the
localization of the bulk gravitational fluctuations on the
three-dimensional brane,  which plays the r\^ole of our world.
This localization property was demonstrated by mapping the problem
of the dynamics of these fluctuations into a one-dimensional
Schr\"odinger eigenvalue problem.

A characteristic feature of such models was the presence of a
massless mode for the
graviton (in agreement with  Lorentz covariance on the brane)
together with a  continuum of  massive Kaluza-Klein (KK)
states on
the four-dimensional world.
These KK modes have
different properties as compared with the factorizable case.
The presence of such KK states
leads to corrections of the
four-dimensional Newton's law; such corrections, however, are
suppressed by quadratic powers of the inverse Planck mass scale, and
hence are unobservable for all practical purposes. 
As a result of the above localization, a solution to the mass
hierarchy emerges in the sense that the weak scale is
generated from a large scale of the order of Planck mass through
an exponential hierarchy, induced by the presence of the warp
factor in the metric of the four-dimensional world.

In the original scenario of \cite{randal99} only 
fields in the gravitational
multiplet have been allowed to propagate in the bulk direction.
The standard model fields were assumed to live exclusively on the 
brane worlds. In the approach of \cite{GW}, however, 
standard model matter fields of mass close to the string scale $M_s$,
which is a natural mass scale in the bulk, were assumed to exist
also in the bulk, where they can propagate normally. 
Within the context of the Randall-Sundrum
space time~\cite{randal99}, then, it was shown that the KK modes
of such bulk fields are characterized by four-dimensional masses
which are also exponentially suppressed, thereby implying a novel 
mass hierarchy. 

In both models~\cite{randal99,GW} one requires 
the existence of a second brane, and  
the distance between  them is supposed to be determined dynamically
by the minimization of a particular potential for the so-called modulus 
(radion) field. 
This minimization is argued to provide a 
stabilization mechanism for the radion~\cite{GW2}.

\section{Quantum Fluctuating (Single) Branes and \\ 
Non-Factorizable Bulk Metrics} 

However, in all the above approaches quantum fluctuations of the branes
have been ignored, given that the latter have been assumed rigid.
For consistency with general relativity, such an assumption 
must be relaxed, if one is dealing with finite-mass solitonic
stringy structures. 
In a previous paper~\cite{lema}, 
we have made an attempt to discuss the incorporation of such effects
in a single $D3$ brane, embedded in a five-dimensional geometry.
A single brane situation may be thought of as the limiting case
where any other brane has run away to infinity (i.e. to a bulk 
distance much larger
than any other distance scale in the problem). 

We have found that the quantum fluctuations of the $D3$ brane,
which can be described by the excitation of (open and closed) 
strings~\cite{kogan96,ellis96,mavro+szabo},
formulated in terms of logarithmic conformal field theory on the 
world-sheet~\cite{lcft},
results in a distortion of the surrounding space 
time~\cite{ellis96,lema}, 
in such a way that 
the five-dimensional 
invariant line element corresponding to the 
induced bulk metric is of non-factorizable form between the 
bulk $z$ and $D3$ brane coordinates $X^I$:
\begin{eqnarray}
&~&ds^2_{f} = e^{-2\sigma (z)}
~g_{IJ}(X^K)dX^I dX^J  -  dz^2,
\nonumber \\
&~&\sigma (z) =-\frac 12 \ln\left( \left|1 -\alpha^2
~z^2\right|\right) \label{bulk}
\end{eqnarray}
In the above formula $\alpha$ denotes  
the width of the momentum fluctuations of the 
$D3$ brane in the bulk direction~\footnote{However, as discussed in 
\cite{lema}, an alternative scenario exists in which $\alpha$ is related
to the momentum transfer during the collision of a macroscopic 
$D1$-brane (string), propagating along the bulk direction, with the $D3$ 
brane world at a certain time $t_0$ 
in the remote past.}. 

It is important to notice that,
as a result of the requirement of convergence of the world-sheet path 
integrals, the four-dimensional (brane) space time, with metric $g_{IJ}$ 
is assumed to be {\it Euclidean}. Passage to a Minkowskian 
space time takes place only at the very end of the computations, 
by means of analytic continuation in the target time. 
In the leading-order approach of ref. \cite{lema}
to the $D3$-brane recoil, the four-dimensional 
metric $g_{IJ}$ is found to be flat.

In the approach of \cite{mavro+szabo},\
the width $\alpha$ can be computed to leading order in a weak string 
coupling
by an appropriate resummation procedure 
over pinched annuli on the world-sheet, with the result 
\begin{equation} 
   \alpha \sim g_s \Delta P 
\label{width1}
\end{equation} 
where $g_s <<1$ is the string coupling, which is assumed weak enough 
so that the world-sheet perturbation theory is valid, and 
$\Delta P$ is the momentum uncertainty (along the bulk direction)
of the $D3$ brane. The latter quantity is bounded from below 
as a result of the stringy 
uncertainty principle, which, to a leading order approximation
for weakly coupled branes, has been derived 
in \cite{mavro+szabo} using logarithmic world-sheet 
conformal field theory methods~\cite{lcft}.

At present the precise value of $\alpha$ is unknown, given that 
this would require knowledge of the wavefunction of the five-dimensional
brane model, which is unknown. 
However, following analogous situations in quantum mechanics,
one may assume a saturation of the uncertainty principle,
which allows $\Delta P$ to be expressed in terms of the minimum
length $\Delta z$. In the case at hand, 
within the world-sheet weakly coupled string model, the latter is found 
to be~\cite{mavro+szabo}:
\begin{equation} 
\Delta z \sim \frac{g_s^{\eta/2}}{M_s}  
\label{width2}
\end{equation} 
where $M_s$ is the bulk string scale. The quantity $\eta$
is associated with modular infinities in the $\sigma$-model 
genus expansion~\cite{mavro+szabo}, and currently cannot 
be computed within the $\sigma$-model perturbation theory.
Its value may be fixed by the requirement that the minimum length
(\ref{width2}) coincides with the one obtained from generic 
kinematical arguments in brane theory~\cite{yoneya},
which implies $\eta=2/3$. Standard string theory would require $\eta=0$. 
In general, for our  
purposes in this paper, one may leave at this stage $\eta$ as an unknown
parameter to be bounded by the current phenomenology.   
 
From (\ref{width2}) and (\ref{width1}), then, one obtains:
\begin{equation} 
   \alpha \sim g_s^{1-\frac{\eta}{2}} M_s 
\label{width}
\end{equation}

The non-factorizable metric (\ref{bulk}) results in 
the following modified Einstein term in the action~\cite{randal99}:
\begin{equation}
S=\int dz e^{-2\sigma (z)} \int d^4 X R^{(4)}[g] + \dots 
\label{einstein}
\end{equation} 
where the superfix $(4)$ denotes four-dimensional quantities, depending 
only on the $X^I$, $I=1,\dots 4$ coordinates.  
Equation (\ref{einstein}) implies that 
the effective 
four-dimensional Planck's constant, for an observer living on the 
$D3$ brane, is:
\begin{equation}
M_P^2 = M_s^3 \int dz e^{-2\sigma (z)} 
\label{fourdim}
\end{equation} 
  
In ref. \cite{lema} we have demonstrated the {\it dynamical} 
formation of horizons, located at $z=\pm 1/\alpha$, which 
effectively restrict the bulk space time inside 
$-1/\alpha < z < 1/\alpha$. 
In the context of the models of ref. \cite{GW} in which matter 
fields are allowed
to propagate in the bulk, the r\^ole of horizons can be understood
simply by considering the positive energy conditions~\cite{wald,horizons} 
for the (Minkowskian version of the) 
space time (\ref{bulk}). It suffices to examine the weakest of these 
conditions, which can be formulated in terms of the stress-energy tensor 
$T_{\mu\nu}$ as follows:
\begin{equation} 
T_{\mu\nu}\zeta^\mu\zeta^\nu \ge 0~, \mu,\nu=1,\dots 5
\label{energy}
\end{equation}
where $\zeta^\mu$ are arbitrary null vectors. 

Using Einstein's equations, one may replace the stress-energy tensor
by the corresponding Einstein tensor in the geometry (\ref{bulk}).      
It is easy to see, then, that the weakest energy condition amounts simply to the requirement: 
\begin{equation} 
              \sigma (z)'' \ge 0
\label{weakest}
\end{equation} 
If the weakest energy condition is violated, all the others are 
violated too. From (\ref{bulk}) we observe~\cite{lema} that
the weakest energy condition (\ref{weakest})
is satisfied {\it only} inside the horizon $-1/\alpha < z < 1/\alpha$. 
This implies that stable 
matter can only exist inside this region of the bulk space 
time~\cite{wald,horizons}.

In view of this result we then define the effective four-dimensional Planck
(\ref{fourdim})  
mass in our model of a single fluctuating D-brane world as
\begin{equation}
M_P = M_s^{3/2} \left(\int_{-1/\alpha}^{1/\alpha} dz 
e^{-2\sigma (z)}\right)^{1/2} \sim 
\frac{2}{\sqrt{3}} M_s \left(\frac{M_s}{\alpha}\right)^{1/2} 
=\frac{2}{\sqrt{3}} \frac{M_s}{g_s^{\frac{2-\eta}{4}}}  
\label{fourdim2}
\end{equation} 
where we used (\ref{width}). 

The analysis of \cite{lema} has demonstrated that  
the space-time (\ref{bulk}) satisfies Einstein's equations, 
provided one includes an {\it excitation} energy term of a `stack' 
of parallel $D3$ branes inside the horizon region, 
described by a four-dimensional term 
\begin{equation} 
  \int dz  \int d^4x \sqrt{g} V(z) 
\label{stuck}
\end{equation}
where the flat-space integral denotes the continuum limit of the sum 
representing the (infinite) 
stack of parallel $D3$ branes in the region $|z| < 1/\alpha$. 
This configuration describes
formally the quantum fluctuations of a single $D3$ brane inside the 
horizon~\cite{lema}. It is important to notice that 
the term (\ref{stuck}) {\it does not} vary with respect to the fifth 
component of the metric, unlike the corresponding bulk cosmological 
``constant'' $\Lambda $ term. The explicit dependence of $V(z)$ on the 
$z$ coordinate in general breaks Lorentz invariance in the bulk,
but this is not unexpected as a result of the presence of the 
fluctuating D--brane. On the other hand, Lorentz invariance remains 
a good symmetry on the $D3$ brane~\footnote{More general models of 
foam, characterized by intersecting branes, or $D3$ branes ``punctured'' by 
D-particle defects also exist~\cite{horizons}, but we shall not deal with such models here.}.

The result for $V(z)$, that solves Einstein's equations in the 
five-dimensional geometry, is:
\begin{eqnarray} 
\frac{V(z)}{12M_s^3}=\frac{1}{2}\frac{\alpha ^2}{(1 + \alpha z)^2}
+\frac{1}{2}\frac{\alpha ^2}{(1 - \alpha z)^2} - 
\frac{\alpha ^2}{|1 + \alpha z|}\delta (1 + \alpha z)
- \frac{\alpha ^2}{|1 - \alpha z|}\delta (1 - \alpha z)
\label{explsol}
\end{eqnarray}
At $z=0$, which was the equilibrium location of the $D3$-brane world,
the excitation energy is positive $V(z)=12\alpha^2$.  
On the other hand, the space-time (\ref{bulk}) is characterised
by a ``vacuum energy'' $\Lambda (z)$
\begin{equation}  
\frac{\Lambda}{24~M_s^3} = -\left( \sigma (z) \right)^2 
\label{cosmconst}
\end{equation}
which {\it vanishes} at $z=0$, and is negative 
for $z \ne 0$ (anti-de-Sitter type bulk). 

As discussed in \cite{adrian+mavro99}, 
the result (\ref{explsol}) and (\ref{cosmconst}) 
would signal {\it supersymmetry obstruction}~\cite{witten95} 
on the original 
$D3$ brane as a result of the quantum recoil 
fluctuations: the excited (fluctuating) state 
of the brane is not supersymmetric, due to a non-vanishing 
excitation energy $V(0)\sim 12\alpha^2$,
whilst 
the ground state (not recoiling branes) could be, as a result of 
the anti-de-Sitter type geometry (\ref{cosmconst}). 
The scale of this supersymmetry obstruction would be set by $\alpha$ then, 
which in view of (\ref{width}), (\ref{fourdim2}), 
would define a {\it new hierarchy} 
for weakly coupled strings $g_s <<1$.

From (\ref{width}), one finds the following hierarchy of scales:
\begin{equation}
    M_s \sim   \left(\frac{3}{4}\alpha ~M_P^2\right)^{1/3} 
\label{hierarchy}
\end{equation}
Indeed, for $M_P \sim 10^{19}$ GeV and  
${\cal O}(1)~{\rm TeV} < \alpha \le {\cal O}(100)$~TeV, 
as appears to be necessary for
a solution to the gauge hierarchy problem on the $D3$ world, 
one would obtain from (\ref{hierarchy}): 
\begin{equation}
  M_s \sim 10^{14}~ {\rm GeV}  
\label{interscale}
\end{equation} 
implying an intermediate string scale.
In this case, the hierarchy (\ref{hierarchy}) 
should be compared with the one encountered 
in a class of
 conventional-field-theory models involving intermediate-scale $M_I$ 
unification, $M_I \sim \sqrt{M_P~m_W}$, where $m_W$ is the electroweak 
scale. 
  
Notice that, in contrast to the model of \cite{randal99} 
involving two parallel branes separated by a distance $r_c$, which sets 
the hierarchy in that case, our model has a {\it single} brane, whose quantum 
(momentum) {\it fluctuations} in the extra dimension are responsible for 
generating a
hierarchy (\ref{hierarchy}) 
between the supersymmetry-breaking-scale and the string (or 
Planck) mass scale (\ref{fourdim2}). 

Notice also, that in case one assumes 
saturation of the uncertainty principle, determining
$\alpha$,
then
from (\ref{width}) and (\ref{fourdim2}) we would obtain: 
\begin{equation} 
g_s^{1-\frac{\eta}{2}} \le 10^{-9}~, 
\label{scales}
\end{equation} 
implying fairly weak string couplings ($g_s \sim 10^{-9}$ for $\eta=0$
(ordinary strings),
or $g_s \sim 10^{-13}$ for $\eta=2/3$ (branes)). Such weak couplings are not 
unusual in string theories dual to ordinary 
string theories~\cite{antoniadis}. 
However, 
as we have explained above, for our purposes here
we prefer to treat $\alpha$ as a phenomenological 
(but, in principle, theoretically calculable) parameter,
to be constrained by the current data.

\section{Localization of Fields on the fluctuating D3 brane world and a novel Mass Hierarchy}

\subsection{Graviton Modes} 

An interesting result of the analysis in \cite{lema} was the demonstration 
of the localization of graviton four-dimensional 
KK modes with masses $m < \sqrt{2}\alpha$ 
on the $D3$ brane, with the simultaneous expulsion of modes with masses higher than $\sqrt{2}\alpha$ on the horizons. 
To show this, we followed
\cite{randal99} and
used the following ansatz for separating variables
$X^I$ and $z$, as far as
(small) quantum fluctuations of
the bulk graviton state
${\hat h}(X^I,z)$ about the background (\ref{bulk})
are concerned:
\begin{equation}
 {\hat h}(X^I,z)=\lambda (z) e^{ip^E_IX^I}
\label{fluct}
\end{equation}
where the notation $p^E_I$ in the momenta on the brane has been
explicitly stated to remind the reader that we are working on a
Euclidean set up for $\{ X^I \}$, and hence for massive
KK excitations, of mass squared $m^2>0$. The on-shell
condition for these modes should then read
\begin{equation}
(p^E_I)^2=-m^2 < 0
\label{onshell}
\end{equation}
The equation for such small fluctuations, can be obtained
by linearizing  Einstein's equations around the
AdS background (\ref{bulk}) 
and choosing an appropriate gauge for the fluctuations
of the metric.  
Upon introducing the ansatz (\ref{fluct}), and using
(\ref{onshell}), the equation becomes a one-dimensional
Schr\"odinger-type eigenvalue equation
for the bulk modes $\lambda (z)$~\cite{randal99,lema}
\begin{equation}
-\lambda '' (z) + \left(4 (\sigma ')^2 - 2 \sigma '' \right)\lambda (z) =
-m^2 e^{2\sigma}\lambda (z)
\label{lambdaeq}
\end{equation}
It is important to note the {\it minus}
sign in front of the mass term on the right-hand-side
of (\ref{lambdaeq}), which is due to the
Euclidean nature
of $X^I$ hyperplane (\ref{onshell}).
As we have explained in \cite{lema}, this plays the important
physical r\^ole of inducing localization of light KK modes 
with masses below a critical mass $m_{cr}=\sqrt{2}\alpha$.

\begin{figure}[t]
\begin{center}
\epsfxsize=3in
\bigskip
\centerline{\epsffile{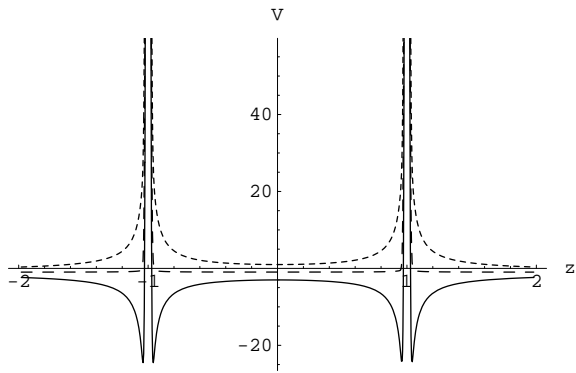}} \caption{\it
Schematic representation of the ``Suspension Bridge''
potential of the equivalent
Schr\"odinger equation 
corresponding to the dynamics of a quantum-fluctuating $D3$ brane
in the bulk direction of a five-dimensional space time. 
The solid curve
corresponds to the massless Kaluza-Klein (KK) four-dimensional mode, 
while the dashed curves represent
the potential for the KK modes with masses 
$m^2\ge m_{cr} = {2} \alpha^2$. 
The wavefunctions of these modes 
are no-longer
localized on the $D3$-brane at $z=0$. The reader is invited
to notice the completely impenetrable potential walls
(denoted in the figure by disjoint lines), which is the result 
of the $D3$-brane bulk quantum-fluctuations (`recoil'). 
\label{fig2}}
\end{center}
\end{figure}

To see this, we recall that (\ref{lambdaeq})  
has the form of a one-dimensional Scrh\"odinger's equation,
with potential:
\begin{equation} 
{\cal V}(z)= - \left(\frac{2\alpha ^2-m^2}{|1 - \alpha^2 z^2 |}
-  \frac{\alpha ^2}{|1 + \alpha z |}\delta (1 + \alpha z)
 - \frac{\alpha ^2}{|1 - \alpha z |} \delta (1-\alpha z)  
\right)
\label{gravpotential} 
\end{equation}
which is depicted in figure~\ref{fig2}.
It is important to notice that this potential has non-trivial 
$\delta$-function (repulsive) boundaries, which imply a dynamical restriction
of the bulk space time inside the region $\alpha^2 z^2 < 1$.

For the region close to the brane
$z = 0$, we may ignore the effect from the boundaries. 
Notice 
that near $z=0$ the potential  
is {\it attractive} 
for
\begin{equation} 
m^2 < 2\alpha ^2 
\label{cond1} 
\end{equation} 
which is the origin of the critical mass for KK 
excitations of massless bulk fields (including gravitons)~\cite{lema}.
The potential (c.f. fig. \ref{fig2})  
starts as a inverted harmonic oscillator potential~\cite{barton} 
around $z=0$, 
but then deviates significantly from it (by
anharmonic terms),
to
end up 
in $\delta$-function opaque (repulsive) walls 
at $z=\pm 1/\alpha$, ${\cal V}(\pm 1/\alpha) \sim \Omega \delta(0)$ 
with 
{\it infinite} opacity $\Omega = \substack{{\rm Lim} \\ z \rightarrow \mp
\frac{1}{\alpha}}\left(\frac{1}{|1 \pm \alpha z|}\right)$.
Therefore the wavefunction should vanish at the boundaries,
thus effectively implying a {\it dynamical} restriction of the 
bulk direction inside the region $z^2\alpha^2<1$. This may be seen as 
another alternative to compactification~\cite{lema}, 
physically 
different from 
the one proposed in \cite{randal99}. 

We now note that away from the boundaries 
the Schr\"odinger equation can be solved analytically.
Requiring $\frac{d\, \lambda }{d\, z}|_{(z=0)}=0$, the  graviton states 
wavefunction is given by
\begin{eqnarray}
\lambda (z)~\sim~{}_2F_{1}\left(-\frac{1}{4}-\frac{3}{4}
                                \left[1-(\frac{2m}{3\alpha})^2\right]^{1/2},
                     -\frac{1}{4}+\frac{3}{4}\left[1-(\frac{2m}{3\alpha})^2\right]^{1/2},
                     ~\frac{1}{2},~(\alpha~z)^2\right)
\label{hypergeometric} 
\end{eqnarray}

{}From the behaviour of the hypergeometric function, we find that
states with mass less than the critical value $m < \sqrt{2} \alpha$ 
have wavefunctions which are peaked 
on the $D3$-brane world at $z=0$. If one, now, includes the $\delta$-function
boundaries, then the resulting wavefunction is localized inside the 
region $\alpha^2 z^2 < 1$~\cite{lema}.  
The massless KK graviton mode ($m=0$) 
is depicted in fig. \ref{fig3}, where the effect of the 
boundaries at $z=\pm 1/\alpha$ in (\ref{gravpotential}) is included. 
For the critical value $m=m_{cr}$ we have
$\lambda _{cr} (z)= {}_2F_{1}(-\frac 12, 0, \frac 12, (\alpha z)^2)=1$, 
corresponding to 
a constant mode (with amplitude normalized to one for convenience),
which is thus not compatible with the vanishing boundary conditions
at $z=\pm 1/\alpha$, imposed by the opaque walls. 
We remind the reader that proper inclusion of the boundary 
effects in the Schr\"odinger equation
(\ref{lambdaeq}) requires numerical treatment. From such an analysis,
we deduce~\cite{lema} that, as a result of 
the $\delta$-functions in the corresponding potential,
states with KK masses 
slightly below the critical value, $m\sim \sqrt{2}\alpha \pm \epsilon$, 
$\epsilon \rightarrow 0^+$,  
are distorted due to the boundary effects at $z=\pm 1/\alpha$, and hence
begin to delocalize. 

On the other hand, the wavefunctions of modes with masses $m > m_{cr}$
are infinitely peaked at the boundaries
$z=\pm 1/\alpha$~\cite{lema}, and hence 
they are unbounded, and could be discarded on physical 
grounds~\footnote{Even if one accepts
unbounded solutions to the original KK mode problem, 
such modes are definitely not localized on our
brane world at $z=0$, and from this point of view are 
of no interest to us here.}.  
The above features are 
easily deduced by the analogy of the corresponding 
Schr\"odinger potential for the case $m>m_{cr}$, near $z=0$, 
with that of an ordinary harmonic oscillator. As can be readily seen,
such a supercritical 
case would correspond to a negative-energy eigenvalue state, 
which does not exist for an oscillator potential  with a positive minimum.

The presence of massive KK modes for the 
graviton on the $D3$ brane world leads~\cite{lema} 
to modifications of Newton's law
on the brane hypersurface
$\{ X^I \}, I=0, \dots 3$ at $z=0$, due to the 
exchange of massive modes with masses up to $\sqrt{2}\alpha$.  
As discussed in \cite{lema}, 
in the case of free boundary conditions
on the horizons $|z|=1/\alpha$, 
there are
(attractive) corrections to Newton's law of $r^{-2}$ scaling, which are
suppressed by a power of $\alpha/M_s$. 
Notice that on account of the hierarchy (\ref{hierarchy}),
for supersymmetry-breaking scales $\alpha \le 100$ TeV, as required for a 
solution to the gauge hierarchy problem, and a string
scale $M_s \sim 10^{14}$ GeV,
the corrections to Newton's law are negligible, consistent with the 
current phenomenology. 

On the other hand, if one imposes periodic boundary conditions 
for the various mode wavefunctions on the horizons, which 
corresponds to {\it compactification}, then the  
KK spectrum is discrete with 
masses $m=\frac{n\alpha}{2}$, $n=0,1,2,\dots $. 
In this case one essentially identifies the two horizons,
and thus one has effectively a two brane scenario.  
In view of (\ref{cond1}), only the first few ($n=0,1,2)$ 
of these KK modes are localized
on the $D3$ brane. This implies that the corresponding corrections 
to Newton's law involve a sum over these few modes:
\begin{equation} 
V_G=-\frac{G_N~m_1~m_2}{r}\left(1 + \sum_{n=1}^{2}
\left(\frac{M_P^2}{c_n^2}\right)e^{-\frac{n}{2}\alpha r}\right)   
\label{newton}
\end{equation}
where $c_n$ are the corresponding couplings, $G_N=1/M_P^2$
is the four-dimensional Newton's constant, and $M_P$ is 
the four-dimensional Planck mass. For our case of $\alpha \le 100$ TeV, 
again such corrections are negligible for distances down 
to submillimeter scales, which is the current experimental accuracy
for such tests.  
  
\subsection{Massive Scalar Fields}

\begin{figure}
\begin{minipage}[b]{8in}
\epsfig{file=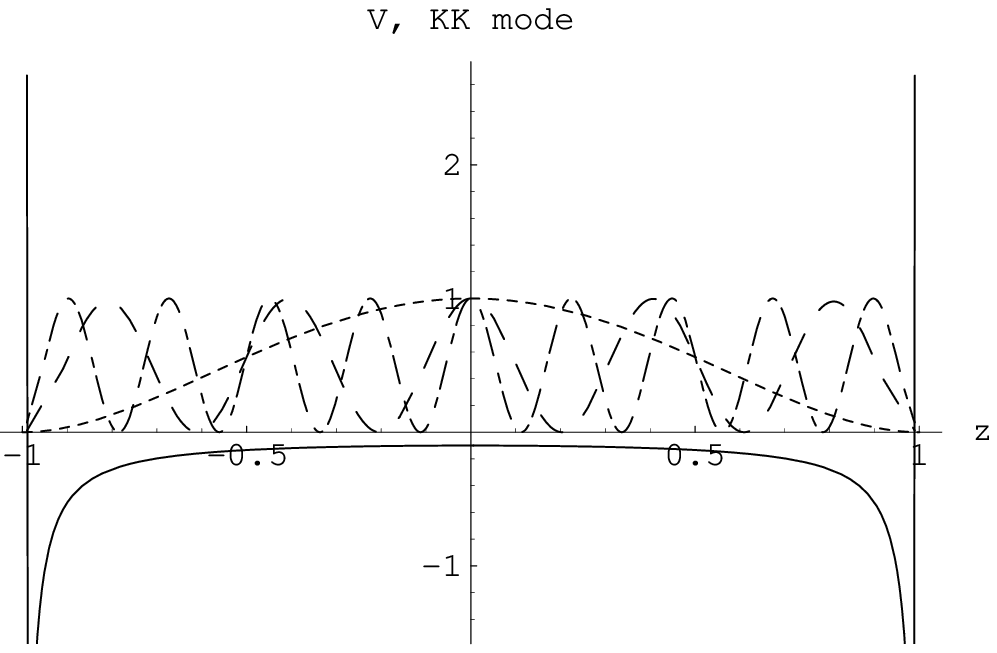,width=2.8in}\hspace*{.3cm}
\epsfig{file=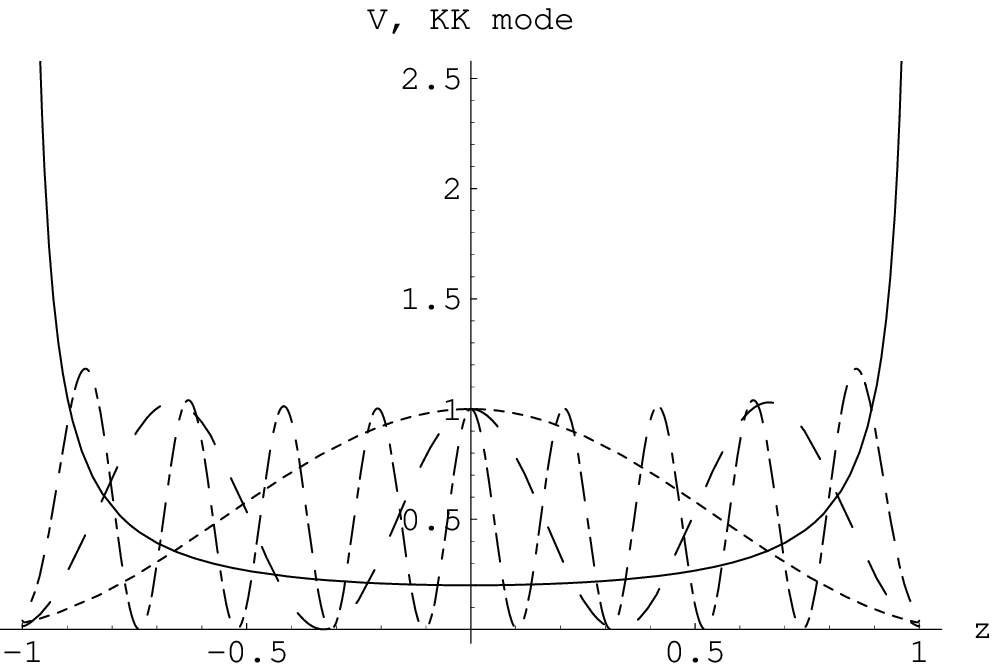,width=2.8in} \hfill
\end{minipage}
\caption{\it Field Localization in Euclidean Space Times. 
The figure on the left shows
the modulus squared of the scalar-field wavefunctions (arbitrarily normalized)
for three light Kaluza-Klein (KK) modes,  with masses below the critical mass
$0\le m< m_{cr}=\sqrt{2}\alpha$. The continuous  line corresponds to the 
potential of the associated Schr\"odinger equation. The dotted curve 
denotes the 
massless $M_5=0$ case (which is formally the same as the graviton case). 
The dashed and dashed-dotted lines, on the other hand, correspond to 
$M_5 \ne 0$ cases. The higher the value of $M_5$, the larger 
the number of nodes of the wavefunction. 
The 
wavefunctions corresponding to these modes 
are peaked 
on the $z=0$ brane world and vanish at the boundaries $z=\pm1/\alpha$. 
The figure on the right shows the modulus squared of 
the wavefunctions of three heavy KK modes 
with masses  $m>m_{cr}=\sqrt{2}\alpha$. The continuous line
denotes again the corresponding Schr\"odinger potential. 
The dotted line corresponds to the case $M_5=m$, and exhibits
localization on the $D3$ brane at $z=0$. The dashed curve
corresponds to the case $M_5 < m$, and has a local peak at $z=0$,
while vanishing at the boundaries. Finally,    
the dashed-dotted curve corresponds to 
the case $M_5 > m$.  
The wavefunctions of these latter modes peak near the boundaries, but 
vanish exactly at the boundaries 
due to the infinite opacity.} 
\label{fig3}
\end{figure}

Next we examine the scenario of ref. \cite{GW}
for matter bulk fields  
within the context of the present model (\ref{bulk}). 
To this end we consider an expansion of a bulk scalar field
in terms of ``modified plane waves'', to take proper account of the 
induced curved metric (\ref{bulk}): 
\begin{equation}
\Phi (X^I,z) = \frac{1}{g^{1/4}} e^{iP_IX^I} \varphi (z) =  
e^{iP_IX^I} e^{2\sigma (z)} \varphi (z)
\label{scfield}
\end{equation} 
Following the approach of 
\cite{GW}, we consider 
the action of a bulk matter scalar field, of five-dimensional 
mass $M_5 $, propagating inside the horizon region 
of the (Euclidean) space time (\ref{bulk}), where such matter can 
live in a stable form due to the energy conditions (\ref{weakest}):
\begin{equation} 
S=\frac{1}{2} \int d^4x \int _{-1/\alpha}^{1/\alpha} \left(-e^{-2\sigma (z)} 
\delta^{IJ} \partial_I \Phi \partial_J \Phi + \Phi \partial_z 
\left(e^{-4\sigma (z)} \partial_z \Phi \right) + M_5^2 e^{-4\sigma (z)}\Phi^2 \right)
\label{action}
\end{equation} 
Upon inserting the decomposition (\ref{scfield}), and 
using the four-dimensional 
on-mass-shell (Euclidean) condition $(P_I^E)^2 = -m_{\rm sc}^2 < 0$, 
we arrive at an one-dimensional Schr\"odinger-like equation for the 
KK mode wavefunctions $\varphi (z)$:
\begin{equation}  
-\varphi '' (z) + \left(4 (\sigma ')^2 - 2 \sigma '' \right)\varphi (z) =
-\left(m_{\rm sc}^2 e^{2\sigma} - M_5^2 \right)\varphi (z)
\label{varphieq}
\end{equation}
which for massless bulk scalar modes $M_5$ is identical to that for the 
graviton KK modes (\ref{lambdaeq}), leading to localization on the $D3$ brane 
of modes lighter than $\sqrt{2}\alpha$.

For massive scalar bulk fields~\footnote{A similar analysis is 
valid for massive vector bulk 
fields.}, with $M_5 \ne 0$, the Schr\"odinger 
equation (\ref{varphieq}) becomes (inside the horizon): 
\begin{equation} 
  -\varphi '' (z) - \left(\frac{2\alpha ^2 - m_{\rm sc}^2}{|1 - \alpha^2 z^2 |}
-  \frac{\alpha ^2}{|1 + \alpha z |}\delta (1 + \alpha z)
 - \frac{\alpha ^2}{|1 - \alpha z |} \delta (1-\alpha z)+ M_5^2 
\right)\varphi  (z) =0~, 
\label{varphifinal}
\end{equation}
There are rigorous theorems 
on the existence of solutions of this equation~\cite{kamke}.
Inside the region of interest, 
$\alpha ^2 z^2 < 1$, away from the boundaries, where the effects
of the $\delta$-functions can be ignored, 
the latter can be expressed as appropriate sums of Bessel and/or
hypergeometric functions. For generic values of the parameters
the solutions are not known in a closed form, but only recursion relations
among the summation coefficients have been provided~\cite{kamke}. 
For the purposes
of our work here, we restrict therefore ourselves to 
a numerical analysis of the solutions, including the boundary terms.

Note that in the massless limit $M_5=0$ the equation 
becomes similar to (\ref{lambdaeq}). 
The potential corresponding to (\ref{varphifinal}) 
has analogous form to 
that in figure \ref{fig2}, the only difference being the positive shift  
by the constant mass term
$M_5^2 \ne 0$. Such terms may be viewed as providing energy-eigenvalue
terms in the corresponding Schr\"odinger equation. 

The potential is attractive, in the interior $\alpha^2 z^2 < 1$,  
only for four-dimensional KK modes of the bulk massive (scalar) fields 
for which (\ref{cond1}) is satisfied. Such KK modes have wavefunctions
which peak at $z=0$. 
The above conclusion is supported by a numerical analysis 
analogous to the one in the graviton case (see fig. 
\ref{fig3}, which depicts the relevant probability densities
for various cases). 
The modes that 
violate (\ref{cond1}), on the other hand, 
correspond to a Schr\"odinger 
potential which is similar to that of an ordinary 
harmonic oscillator near $z=0$ (c.f. fig. \ref{fig2}), 
but deviates significantly from it
by anharmonic terms as one goes to higher values of $z$ to end 
up in two infinitely opaque walls at $z=\pm1/\alpha$.

The energy spectrum is {\it discrete}, given that 
the situation is equivalent to that of a particle
{\it bound} inside a potential well with infinitely opaque 
walls~\footnote{We stress that this is an exclusive feature of the
completely impenetrable potential-wall terms 
$\frac{1}{1\pm \alpha z}\delta (1 \pm\alpha z)$,
induced by the $D3$-brane recoil. If only 
$\delta$-function terms 
were present, as in the conventional $D$-brane-inspired models of 
\cite{randal99}, where recoil is ignored, 
then there would be a finite possibility of wall penetration, in which case
one would also have continuous energy spectra, depending 
on the value of the energy.}. We note that 
the discreteness of the spectrum
is a generic feature of the recoil-induced space-times  
with anti-de-Sitter like bulk~\cite{mavro+winst2000},
such as (\ref{bulk}). 
One therefore may expect nodes in the wavefunction, 
whose number depends on the
value of the energy $M_5^2$ (Fubini's theorem~\cite{messiah}). 
This is confirmed by a numerical analysis of the equation (\ref{varphifinal}).
The discreteness of the energy spectrum is interesting in that 
it implies discrete allowed values for the bulk mass $M_5^2$ in terms 
of the quantity $|m^2 - 2\alpha^2|$, and hence quantization of 
$m$ for given $M_5,\alpha$.

Unfortunately,
for generic values of the parameters the corresponding spectrum cannot be 
obtained analytically. 
However, one may obtain approximate
solutions in certain cases, which capture the qualitative features. 
For instance, in the massless case $M_5=0$, discussed in the previous 
subsection, 
away from the boundaries,
the Schr\"odinger equation becomes hypergeometric (\ref{hypergeometric}). 
In the massive case, $M_5 \ne 0$, and for modes $m^2 < 2\alpha^2$ 
one may consider the Schr\"odinger equation (\ref{varphifinal}) 
for $\alpha^2 z^2 << 1 $, and expand the binomial $(1-\alpha^2z^2)^{-1}$
up to the first power of $(\alpha z)^2$. In this case the problem is 
equivalent to 
a Schr\"odinger equation with an inverted harmonic oscillator 
potential~\cite{barton}. By changing variables to 
$z \rightarrow x \equiv \sqrt{2}\alpha z (2-m^2/\alpha^2)^{1/4}$, 
and defining 
$a\equiv -(M_5^2 + 2\alpha^2 -m^2)/(2\alpha\sqrt{2\alpha^2 -m^2})<0$,
the Schr\"odinger equation acquires the form:
\begin{equation}
\frac{d^2}{dx^2}\varphi (x) + \left(\frac{1}{4}x^2 - a\right)\varphi (x)=0 
\label{cylpar}
\end{equation}
The eigenfunctions in this case 
are appropriate combinations of parabolic cylinder functions
$W(a,x)$,  
which are actually oscillatory 
for all $x$ (since $a<0$)~\cite{abram}. 
The requirement that the solution 
has a maximal probability density 
on the $D3$ brane at $z=0$ (so as to imply some sort of 
localization) implies that only even functions of $x$ should be considered. 
The appropriate combination is $\varphi (x)=W(a,x)+W(a,-x)$, in the notation 
of \cite{abram}.  
In the regime of interest to us here
$x^2 << -a$ $(\alpha^2 z^2 <<1$), 
which implies the asymptotic form:  
\begin{eqnarray}
&~&\varphi = W(a,x)+W(a,-x) \sim 2W(a,0)e^{v_r(x)}{\rm cos}\left(\sqrt{-a}x+ u_i(x)\right)~, \nonumber \\
&~&W(a,0)=2^{-3/4}\left|\frac{\Gamma(\frac{1}{4}+\frac{1}{2}ia)}
{\Gamma(\frac{3}{4}+\frac{1}{2}ia)}\right|^{1/2}~, \nonumber \\
&~&v_r(x) \sim -\frac{(x/2)^2}{(2\sqrt{-a})^2} + \frac{2(x/2)^4}{(2\sqrt{-a})^4}
-\frac{9(x/2)^2 +(16/3)(x/2)^6}{(2\sqrt{-a})^6} + \dots~, \nonumber \\
&~& u_i(x) \sim \frac{(2/3)(x/2)^3}{2\sqrt{-a}} - 
\frac{(x/2) + (2/5)(x/2)^5 }{(2\sqrt{-a})^3} + \frac{(16/3)(x/2)^3 
+ (4/7)(x/2)^7}{(2\sqrt{-a})^5} - \dots~. 
\label{cosform}
\end{eqnarray}
where $v_r(x)$ is an even function of $x$,  
and $u_i(x)$ is an odd function of $x$. 
The requirement that the solution  
vanishes at the opaque boundaries $z = \pm 1/\alpha$~\footnote{Notice that
the boundary region $z=\pm1/\alpha$ 
lies outside the regime of validity of the 
inverted-oscillator 
approximation. Nevertheless, for the qualitative description 
of the discrete spectrum, 
it is sufficient to impose the above boundary 
condition to the solutions of the equation (\ref{cylpar}).}, 
implies the vanishing of the argument of the cosine in (\ref{cosform}),
and thus 
a discrete $a$-, and hence KK $m$-, spectrum, for given $M_5,\alpha$:
\begin{equation} 
\sqrt{\frac{M_5^2 + 2\alpha^2 - m^2}{\alpha^2}} + 
u_i(\sqrt{2}(2-m^2/\alpha^2)^{1/4}) \simeq (n + \frac{1}{2})\pi~, \qquad n=0,1,2,\dots 
\label{energydiscr} 
\end{equation} 
under the constraint $0<m^2<2\alpha^2$,
which restricts severely the allowed spectrum of KK modes. 
For instance, for $M_5 >> \alpha$,
this constraint, in conjunction with (\ref{energydiscr}), 
implies that the allowed KK modes have $n \sim M_5/\alpha >> 1$.  
Of course, it is understood that the above approximate analysis
is strictly valid 
for $\alpha ^2 z^2 << 1$,
and should only be considered as indicative for the real case. 
There is one limiting case, though, 
where the above results are valid throughout
the region $0<\alpha^2 z^2 <1$, 
and this is when one is dealing with near-criticality
KK modes with mass $m \rightarrow \sqrt{2}\alpha -\epsilon, \epsilon \rightarrow 0^+$, in the case $M_5 \ne 0$. 
In this case, from (\ref{energydiscr}), 
we observe that there is only a discrete allowed set of bulk masses 
$M_5 =(n+\frac{1}{2})\pi \alpha$, $n=0,1,2 \dots$, which is compatible
with the vanishing $\varphi$ 
boundary conditions at $z=\pm1/\alpha$ set by the opaque walls. 
The above qualitative
results are supported by a numerical analysis of the complete problem
including properly the boundary effects
(c.f. fig. \ref{fig3} for   
the modulus-squared of the wavefunctions of 
subcritical KK modes, 
which peak on the $D3$ brane at $z=0$). 

A similar approximate eigenvalue analysis may be applied 
to KK modes with masses  
$m^2 > 2\alpha^2$. In this case, for $|\alpha z| << 1$,  
one may deduce the approximate energy-eigenvalue spectrum $E_n$
by analogy with the harmonic oscillator case: 
$E_n \equiv M_5^2 = m^2 - 2\alpha^2 + 
\alpha \sqrt{2m^2 - 4\alpha^2}\left(n + \frac{1}{2}\right)~$, 
$n=0,1,2, \dots $. 
Inverting this relation with respect to $m$ one obtains:
\begin{equation} 
m^2 = 2\alpha^2 + \frac{1}{2}\left[ -\alpha (n + \frac{1}{2}) 
+ \sqrt{2M_5^2 + \alpha^2(n+\frac{1}{2})^2}\right]^2~,~n=0,1,2, \dots  
\label{discretespect2}
\end{equation} 
We notice that the analogy with the harmonic oscillator system, 
with a positive minimum
in the potential,
implies the following restriction on KK modes (positivity of the 
``oscillator energy eigenvalue''):
\begin{equation} 
  m^2 < M_5^2 + 2\alpha ^2 
\label{restriction}
\end{equation} 
It is straigthforward to determine the range of $m^2 (>2\alpha^2)$
for which (\ref{restriction}) is valid
together with $M_5^2 \le m^2$: 
\begin{equation} 
2\alpha^2 < m^2 < 2\alpha^2 + 
\frac{8\alpha^2}{(2n+1)^2}~, \qquad n=0,1,2, \dots 
\label{M5lem}
\end{equation}  
Outside this region, $M_5^2 > m^2$.   

Note that in standard string models 
one allows only the (massless) gravitational string multiplet to 
propagate in the bulk~\cite{extra}, which implies $M_5=0$. 
In view of the harmonic oscillator analogue, such a 
state does not exist (in a bounded form~\footnote{To avoid 
confusion we stress, once again, 
that the
acceptable solutions corresponding to the discrete set 
of `energy' eigenvalues in the analog Schr\"odinger 
problem are only 
the ones that are bounded, i.e. vanish at the boundaries $z=\pm1/\alpha$.
This should always be taken properly into account in any numerical 
treatment (see fig. \ref{fig3}).})
in the case $m^2 > 2\alpha^2$. 
A numerical analysis, supported analytically by 
the harmonic oscillator analogy, indicates that 
bounded KK modes 
with masses $\sqrt{2}\alpha < m < M_5$, have 
wavefunctions that peak near (but not exactly at) the boundaries, 
symmetrically about the origin, 
but then vanish at the boundaries, due to the infinite
opacity of the walls. 
However, for the case $M_5=m$, one obtains a discrete set of 
allowed values of $m > \sqrt{2}\alpha$ which, as demonstrated 
in fig. \ref{fig3},  appears to 
exhibit localization on the $D3$ brane at $z=0$. 
This is a feature that persists in the Minkowskian treatment as we shall 
discuss in the next subsection. 
Finally, the bounded wavefunctions for KK modes with $m > M_5$, under the 
constraint (\ref{restriction}), 
appear to have local peaks at the $z=0$ but not strong localization there
(see fig. \ref{fig3}). 
Again this feature survives the analytic continuation
procedure to Minkowskian times. 

Our analysis
above demonstrates, therefore, that in the scenario of \cite{GW},
where matter fields are also allowed to propagate in the bulk, 
the incorporation of stringy `recoil' (fluctuation) 
bulk effects of the $D3$ brane will imply at most a discrete $M_5$ mass 
spectrum, or, inversely, 
a discrete set of KK masses $m$ for given $M_5,\alpha$ 
(c.f. (\ref{energydiscr}),(\ref{discretespect2})).

We now make an important comment. 
In the stringy problem a 
natural mass scale in the bulk~\cite{GW} is the string scale $M_5 \sim M_s$.
In that case, as we have seen, a natural upper bound on $\alpha$ is 
a few hundreds TeV~\cite{adrian+mavro99} 
in order to provide a satisfactory solution
to the gauge hierarchy problem in supersymmetric models. Hence, 
one has a strong 
hierarchy of scales:   
$M_s \sim 10^{14}~{\rm GeV} >> \alpha \sim {\cal O}(100)~{\rm TeV}$. 
It therefore follows 
that bounded KK modes with masses $m \ge M_5$ 
do not exist in such models,  
in view of the restriction (\ref{M5lem}), which-we stress- 
is an exclusive feature of the Euclidean treatment. 
Thus, one has effectively strong localization 
on the $D3$ brane at $z=0$ 
only 
of light KK modes of matter fields with masses 
$m \le \sqrt{2}\alpha$,
as in the graviton case~\cite{lema}. 
We shall come back to this issue in the next subsection.

\subsection{Analytic Continuation to Minkowskian space times and Field Localization}

\begin{figure}
\begin{minipage}[b]{2in}
\epsfig{file=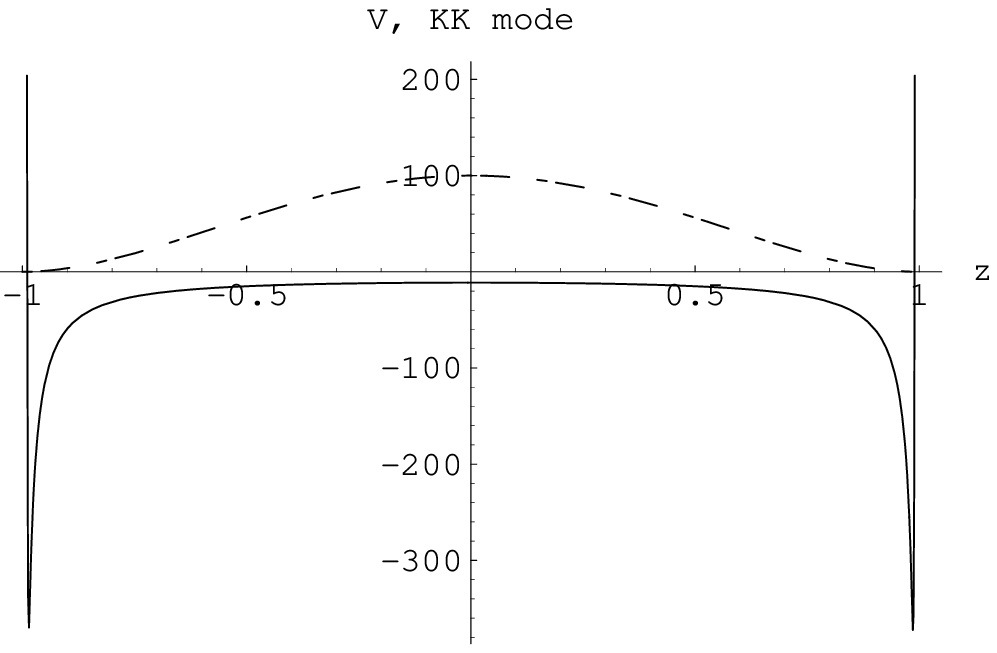,width=2.8in}\hspace*{.3cm}
\epsfig{file=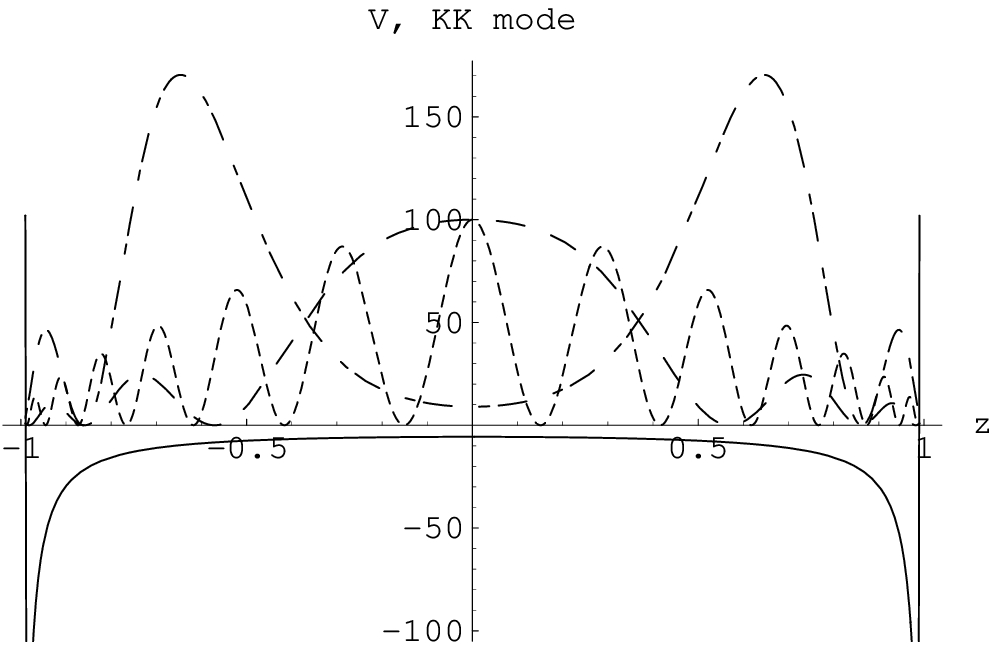,width=2.8in} \hfill
\end{minipage}
\caption{\it 
Field Localization in  the case of {\it Minkowskian} time. 
The figure on the left shows
the modulus squared of the massless scalar-field 
wavefunction (or, equivalently, that of a graviton
mode), arbitrarily normalized, 
for Kaluza-Klein (KK) modes with masses below the critical mass
$0\le m< m_{cr}=\sqrt{2}\alpha$ (dashed dotted curve). The
wavefunctions corresponding to these modes 
are peaked 
on the $z=0$ brane world and vanish 
at the opaque boundaries. The figure on the right shows
three KK modes of a {\it massive} bulk scalar field of mass $M_5$,  
with masses $m^2 > m^2_{cr}$:  
one with mass  $ m^2 \sim M_5^2$ (dashed line),
which exhibits strong localization on the $D3$ brane at $z=0$, 
one mode with mass $m^2 < M_5^2$,  
(dashed dotted curve), and one mode with mass such that 
$m^2 > M_5^2$ (dotted curve). 
The Solid curves in both figures correspond to
the corresponding potential of the equivalent Schr\"odinger equation.}
\label{fig4}
\end{figure}

In this subsection we discuss the above-described field localization 
after analytic continuation back to (physical) Minkowski target time.
This procedure implies that in all the relevant formulas 
of the previous subsection there will be now 
a change in sign of mass-squared terms in the potential
of the respective Schr\'odinger equations (\ref{lambdaeq}),(\ref{varphieq}):
\begin{equation} 
m^2 \rightarrow -m^2~, \qquad M_5^2 \rightarrow -M_5^2 
\label{minkowskiancont} 
\end{equation} 

In such a case, viewing again the $-M_5^2$ terms as energy-eigenvalue terms
in the corresponding Schr\"odinger equations, 
we observe that the potential at $z=0$ is attractive now for any value of $m$.
Nevertheless, one obtains {\it qualitatively} the same localization behaviour
for the light KK modes $m < \sqrt{2}\alpha$ in the massless case $M_5=0$ 
as 
in the corresponding Euclidean 
case of the previous section, as can be confirmed by a numerical analysis 
of the respective Schr\"odinger equation (see fig. \ref{fig4}
where the respective probability densities are plotted). 
This is a nice consistency check of the Euclidean approach, which, as explained in 
\cite{lema}, was necessitated by the requirement of 
convergence of the world-sheet
path integrals employed in the formalism. 

The case $M_5 \ne 0$ of the Minkowskian treatment
requires particular attention.  
As in the subcritical Euclidean case ($m^2 < 2\alpha^2$), 
but here for all $m$, 
the situation  
is equivalent to that of a particle {\it bound} 
inside a potential well (see fig. \ref{fig4}), which 
starts as an inverted harmonic oscillator potential~\cite{barton} 
around $z=0$, 
but then deviates significantly from it (by anharmonic terms), to
end up 
in $\delta$-function opaque (repulsive) walls 
at $z=\pm 1/\alpha$, 
with 
{\it infinite} opacity.
Therefore the wavefunction should vanish at the boundaries.
The energy spectrum is again {\it discrete}, and   
one has nodes in the wavefunction, 
whose number depends on the
value of the energy $-M_5^2$~\cite{messiah}. 
This is indeed confirmed by a numerical analysis of the 
Minkowskian-time version of the equation (\ref{varphifinal})
(c.f. (\ref{minkowskiancont})).

It should be noticed that the 
precise behaviour of the solutions in the 
allowed region 
of the bulk space, $\alpha^2 z^2 < 1$, 
depends crucially on the relative magnitude of the scales
$m$ and $M_5$ (under the appropriate constraints 
implied by the discreteness of the allowed energy spectrum). 
Specifically, for masses $M_5 > m$,  
due to the shape of the potential, 
which dives down near the boundaries,
before the $\delta$-function terms take over, 
the wavefunctions peak
near the boundaries, symmetrically about the origin, 
as in 
the corresponding 
Euclidean case.  
They of course vanish at the 
boundaries, due to the infinite
opacity of the walls (see fig. \ref{fig4}). 
The distance between the above-mentioned peaks 
depends on the difference 
$M_5^2 -m^2$. The two peaks meet when $M_5^2 \sim m^2$,  
leading to a strong {\it localization} 
on the $D3$ brane at $z=0$ of such KK modes  
(see fig. \ref{fig4}). 
As we have seen, 
this last feature also characterizes the Euclidean treatment. 

Finally, in the case $m^2 > M_5^2 $, 
the wavefunction of the corresponding mode 
resembles that of a free 
particle, included in an envelope which peaks at $z=0$ 
and vanishes at $z=\pm 1/\alpha$
(see fig. \ref{fig4}). Thus, such KK modes  
are {\it not strongly 
localized } 
on the $D3$ brane at $z=0$, in contrast to the $M_5=0, m<\sqrt{2}\alpha$,
or $M_5 \sim m > \sqrt{2}\alpha$ cases. Similar features
have been observed for the corresponding case 
in the Euclidean treatment (c.f. fig. \ref{fig3}).

We should note, however, that 
in the stringy case the latter modes ($m > M_5$) are superheavy,
given that 
$M_5 \sim M_s \sim 10^{14}$ GeV.
The superheavy KK modes that are allowed then in the 
physical low-energy spectrum are those with masses $M_s < m < M_P$ 
with the Planck Mass $M_P\sim 10^{19}$ GeV, 
being viewed as a low-energy physics 
ultraviolet cut-off.  
However, consistency with the Euclidean 
case, for which the constraint (\ref{restriction}) is valid, 
implies that $M_0^2 \equiv M_s^2 + 2\alpha^2$ is the highest
allowed 
mass scale for the superheavy KK modes in the string-inspired model.
In addition, as we mentioned at the end of the previous subsection,
if one restricts the supersymmetry obstruction 
scale $\alpha $ to a few hundreds of TeV at most, 
motivated by a solution to the gauge hierarchy 
problem~\cite{adrian+mavro99}, 
then 
KK modes with $m \ge M_5$ do not exist, and one effectively 
obtains localization on the $D3$ brane of light KK modes only, with masses
$m \le \sqrt{2}\alpha$.

\section{Brane Quantum Fluctuations in the Randall-\\Sundrum Model} 

So far we have examined the case of a single brane, representing our world.
An interesting issue is to consider the above model~\cite{lema}
in the context of the Randall-Sundrum model~\cite{randal99} involving 
two branes, which are assumed to be stabilized, e.g. through the 
mechanism of \cite{GW2}. 
In this respect, we should stress that the non-factorizable metrics
induced by quantum fluctuations of the individual branes
in the Randall-Sundrum (RS) scenario will induce effects that are suppressed
by powers of the string coupling $g_s$. 
This is due to the fact that the recoil fluctuations of the individual 
branes are string loop effects, as explained in detail in \cite{ellis96}. 

As a result, the induced metric (\ref{bulk}) will appear as a
{\it small} quantum correction to classical 
RS scenaria involving two parallel branes
separated by a finite distance $r_0$, one of which represents the 
observable world. Thus, the classical 
Einstein's equations are no longer expected to 
be satisfied for the system of the two (parallel) fluctuating branes.
In this sense, our quantum fluctuations play the r\^ole
of inducing a scenario that is similar to the one suggested in 
\cite{GRS}. Below we shall present it in a version~\cite{witten2000} 
most suitable for our case here. 

Let us consider the non-factorisable RS space-time metric, with a small 
perturbation $\varepsilon$ representing our quantum-recoil effects: 
\begin{equation} 
ds^2 = \left(e^{-2k_0 z} + \varepsilon |1-\alpha ^2~z^2 |\right) 
\sum_{I=1}^{4} dX_I dX^I - dz^2  
\label{perturbedRS}
\end{equation} 
where $k_0$ is the parameter of the RS solution, whilst $\varepsilon$
is assumed proportional to an appropriate power of the string coupling 
$g_s <<1$, as a result of the {\it quantum } nature 
of the recoil-induced metric. 
Note that, in the case of isolated $D3$ branes examined above,
such proportionality constant may be absorbed in normalization of the 
coordinates, but here, in view of the difference in the relative 
strength of the 
effect as compared to the {\it classical} RS solution, such 
factors must be kept. 

In the RS scenario the bulk geometry has been restricted to the positive $z>0$
portion of the anti-de-Sitter space. One may continue
to the negative $z$ axis, in which case in view of the 
AdS/CFT correspondence~\cite{maldacena}, the five-dimensional 
gravity theory would be 
equivalent to a four-dimensional conformal field theory 
located in the boundary of the AdS space time at $z=-\infty$.
The truncated version of this space time in the RS scenario 
($z>0$) does indeed have 
dynamical four dimensional gravity.
In a similar manner, the quantum-recoil induced space time (\ref{bulk}),
which also arises originally in a truncated version for $z>0$,
due to the recoil formalism, 
and then is analytically continued in the form (\ref{bulk}),    
as 
explained in \cite{lema}, also gets dynamical gravity localized 
on the brane in the way described above. 

In the modification of the RS scenario presented in ref. \cite{GRS},
and elaborated further by Witten in \cite{witten2000}, 
which we adopt here,  
the quantum fluctuations would be unimportant
as long as $e^{-k_0z} >> \varepsilon$, 
and one would get 
the RS scenario. 
This would imply that, for 
four-dimensional distances not so large 
relative to the separation between the two branes, 
the world would look effectively four dimensional,
due to the field localization on our brane world characterizing  
the conventional RS 
scenario~\cite{randal99}. 
In our framework, this would mean that the quantum fluctuations
of the $D3$ brane world in the bulk would be unimportant 
($\varepsilon \rightarrow 0$,
or -in our case- very weak string couplings $g_s \rightarrow 0$). 

On the other hand, when one looks at four-dimensional scales sufficiently large
as compared to the separation between the branes, 
when the quantum fluctuations
of the individual branes become important, then 
one would expect the space time to  
become effectively five dimensional.   
However, 
in our quantum-fluctuating (`recoil') 
case, the positive energy condition, required probably for stability
of the space time~\cite{witten2000}, is valid strictly inside the 
dynamical horizons of extent $z\sim 1/\alpha$ on each side of 
a fluctuating brane. The case in which the quantum fluctuations 
become important would require relatively large $\varepsilon$, or, 
equivalently, 
strong string couplings $g_s$, and hence large $\alpha$ 
(c.f. (\ref{width1})). This, in turn, 
would imply that stable space times are localized within tiny horizons 
around each individual brane, where the light matter fields 
are also localized, as we have seen above.  
Hence, this would prevent a possible detection of the 
fifth dimension by going to sufficiently large four-dimensional distances.  
We consider this an interesting feature of our 
quantum-fluctuating brane approach, which should 
be explored further in the context of higher-dimensional bulk space times.

\section{Conclusions} 

Above we have discussed some interesting 
physical features of a brane-world model for the 
dynamical generation of a non-factorizable
geometry~\cite{lema} through quantum (recoil) effects,
representing position and momentum fluctuations of the brane.
It is important to stress that 
the scale of momentum fluctuations of the brane world,
which are related to the position fluctuations through some 
stringy uncertainty relations~\cite{mavro+szabo},  
imply a mass hierarchy on the four-dimensional world. 
This happens  
without the need for 
implementing a second
brane, and thus stabilization mechanisms,
which is the case of the 
original Randall-Sundrum scenario~\cite{randal99}.  

If, on the other hand, 
one starts from the two-brane scenario of \cite{randal99} 
and \cite{GW}, then the recoil metric
(\ref{bulk}) may be viewed as a small perturbation, expressing 
quantum string corrections. However, even in that case, 
the induced space time appears to be stable (for 
the existence of matter) only in a 
small region around the fluctuating defect, thus preventing 
a possible detection of the fifth geometry by 
going to very large four-dimensional distances as claimed in \cite{GRS},
where the brane fluctuations are ignored.

An interesting feature of our approach is that, as a result of the 
bulk quantum fluctuations (`recoil') of the $D3$ brane, the mass spectrum 
of the four-dimensional 
KK modes is {\it discrete}, which 
seems to be a generic 
feature of recoil-induced space times with anti-de-Sitter 
like bulk~\cite{mavro+winst2000}. This comes about as a 
consequence of the specific form 
of the potential of the equivalent Schr\"odinger problem, 
which is characterised by the presence of completely impenetrable 
potential walls, effectively binding the analogue 
quantum-mechanical `particle' 
inside the region $\alpha^2 z^2<1$. This implies an interesting 
phenomenology. In particular,
in the context of  string-inspired models, 
one finds that, as is the case with the 
light KK modes with masses $m^2 < 2\alpha^2$,  
one also has a strong localization on the $D3$ brane 
at $z=0$ of intermediate-scale massive 
modes $m \sim M_5 \sim M_s \sim 10^{14}$ GeV. Such heavy matter particles
may have interesting phenomenological, as well as cosmological 
implications (e.g. in the cold Dark-Matter problem etc). 

It must be noted, however, that such modes
are absent from the Euclidean Time spectrum of 
string-inspired models in which $\alpha \le 100~{\rm TeV}  
<< M_s\sim 10^{14}~{\rm GeV}$, in order to provide a solution
to the conventional gauge-hierarchy problem. As stressed repeatedly
in the text, 
the Euclidean formalism is necessitated by the 
requirement of convergence of the 
world-sheet approach to the quantum fluctuating (`recoiling') 
branes~\cite{lema}.
Hence, 
it seems to us, that consistency of the Minkowskian 
treatment with such Euclidean formalisms needs to be sought. Under this 
restriction, then, heavy KK states of mass 
$m \ge M_5$ do not exist in a bounded form, and hence 
only light KK modes of
bulk graviton/matter fields, with four-dimensional 
masses  $m \le \sqrt{2}\alpha$,
are localized on the $D3$ brane world at $z=0$ in such string-inspired models. 
However, as we stressed previously, 
our analysis in the present article is more general. 

There are many features of the model that need to be worked out
before definite conclusions are reached on realistic phenomenological issues.
The most important of them is the rigorous 
implementation of the above formalism 
of recoiling (quantum fluctuating) D-branes 
in target-space supersymmetric theories. We have seen that the 
recoil will most likely obstruct supersymmetry~\cite{adrian+mavro99},
but the detailed mechanism of this obstruction and the associated
mass splittings, that would affect 
the phenomenology of the model, 
have still to be worked out.
Nevertheless, we feel that the features already obtained from 
this, rather generic, model of quantum fluctuating branes deserve 
further studies, especially in connection with possible experimental 
signatures that could 
be testable in future colliders or astrophysical experiments. 

\section*{Acknowledgements}

This work is partially supported by the European Union 
(contract ref. HPRN-CT-2000-00152). 
Parts of it have been presented (by N.E.M.) at the HEP2000
International Workshop of the Hellenic Society for High Energy Physics, 
Ioannina, Greece, April 19-23 2000. We thank the organizers
of that meeting for their interest in our work. 
N.E.M. also wishes to thank H. Hofer for his interest and support.

\end{document}